\documentclass[10pt,letterpaper,twocolumn]{article} %% two column, final layout two ccolumn quantum coherence therotetical inve
\usepackage{ol2}
\usepackage[draft]{hyperref}
\usepackage{amsmath}

\begin{document}

\twocolumn[ %% activate for two-column option

\title{Resolution of objects within subwavelength range by using the near field of a dipole}

%% For REVTeX it is possible to automate superscript and e-mail callouts with the superscriptaddress option; see REVTeX4 documentation.

\author{Aziz Kolk\i ran,$^{1,2,*}$ and G. S. Agarwal$^3$}

\address{
$^1$Dept. of Electrical and Electronics Engineering, \.{I}zmir Katip \c{C}elebi University,35620 \c{C}i\v{g}li, \.{I}zmir, Turkey\\
$^2$Dept. of Electrical and Electronics Engineering, Gediz University, 35665 Menemen, \.{I}zmir, Turkey\\
$^3$Department of Physics, Oklahoma State University, OK, Stillwater 74078, USA \\
$^*$Corresponding author: aziz.kolkiran@gmail.com}

\begin{abstract}We analyze the far field resolution of apertures which are illuminated by a point
dipole located at subwavelength distances. It is well known that
radiation emitted by a localized source can be considered a
combination of travelling and evanescent waves, when represented
by the angular spectrum method. The evanescent wave part of the
source can be converted to propagating waves by diffraction at the
aperture thereby it contributes to the far field detection.
Therefore one can expect an increase in the resolution of objects.
We present explicit calculations showing that the resolution at
the far zone is improved by decreasing the source-aperture
distance. We also utilize the resolution enhancement by the near
field of a dipole to resolve two closely located apertures. The
results show that without the near field (evanescent field) the
apertures are not resolved whereas with the near field of the
dipole the far zone intensity distribution shows improved
resolution. This method eliminates the requirements of near-field
techniques such as controlling and scanning closely located tip
detectors.
\end{abstract}


 ] %% activate for two-column option

\noindent It is well known \cite{Vigoureux1,Wolf,Agarwal} that the
use of evanescent waves can overcome the Rayleigh resolution limit
because they contain high spatial frequencies and yield
information about the smallest sub-wavelength details. However,
evanescent waves are non-radiative and they diminish
exponentially. They can be collected and converted to propagating
waves by small detectors of subwavelength dimensions
\cite{Paesler,Betzig,Oshikane}. Such a mechanism is called
near-field detection and requires nanometric positioning to
samples \cite{Novotny}. It is analogous to tunneling microscopy by
a tip detector. On the other hand, advances in meta-materials has
enabled new class of lenses which compensate for the evanescent
loss and thus restoring an image below the diffraction limit
\cite{Pendry,Fang}.

In the present work we propose a different approach to resolution
enhancement. We consider an infinitesimal dipole located at a
subwavelength distance to an aperture. Radiation from a localized
source can be considered as a diverging spherical wave
$\exp(ikr)/r$ with suppressed $\exp(-i\omega t)$ time dependence,
where $k=\omega/c=2\pi/\lambda$. The angular spectrum of such a
wave has an infinite spectrum of spatial frequencies $k_x$, $k_y$.
On propagation from source to the image, high-frequency components
or evanescent part ($k_x^2+k_y^2>k^2$) are filtered out.  Only low
frequency waves ($k_x^2+k_y^2<k^2$) can travel to the far zone.
From the usual theory of diffraction, a small object illuminated
with a propagating wave generates diffracted evanescent modes, and
conversely, by applying the reciprocity theorem, a small object
located in an evanescent field converts part of this field into
propagating waves \cite{Guerra1,Guerra3}.  An aperture located in
the vicinity of a point source diffracts the evanescent part into
radiation. Thus, the subwavelength details of the aperture is to
be carried to the far zone and the method do not require
point-by-point, time-consuming scanning detection. We remark here
that there are various physical systems to implement our proposal.
The current technology enables an excited atom, molecule or a
quantum dot to be used as a dipole
\cite{Gerber,Aharonovich,Rogobete}. A recent experiment
\cite{Sandoghdar} showed that a semiconductor quantum dot and a
near-field coupling nano-antenna could be employed to produce and
detect single photons with almost perfect collection efficiency.

\begin{figure}[htb]
\vspace{-.5cm}\scalebox{1.1}{\includegraphics[width=7.0cm,trim=-.2in
1.5in .1in 1.0in,clip=true]{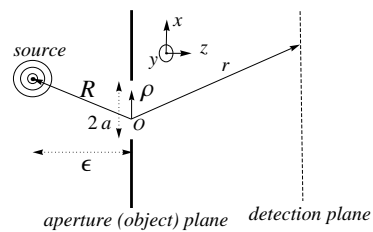}}\vspace{-.6cm}
\caption{Illustration of the coordinates for diffraction of a
dipole field by a circular aperture.}\vspace{-.6cm}\label{Fig1}
\end{figure}

\begin{figure}[h]
\centering {\includegraphics[width=200pt]{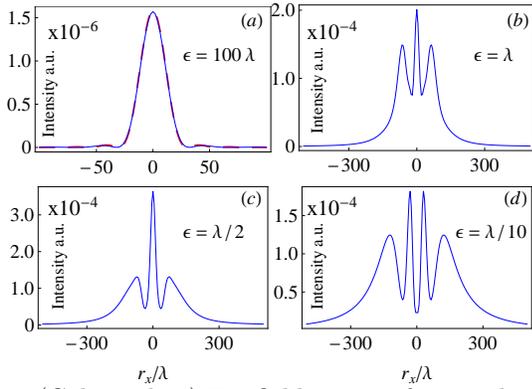}}
\vspace{-.6cm}\caption{(Color online) Far field image of a
circular aperture of radius $a=2\lambda$. In ($a$) the dipole is
located in the far zone and the intensity pattern is an \emph{Airy
disk} (solid line). To check the validity of Fraunhofer
approximations we also show the result (dashed line) of Kirchoff
integral given in Eq. (\ref{Kirchoff}). Figures ($b$) to ($d$)
show intensities with decreasing dipole-aperture distance,
$\epsilon$. $r_z$ is $100\lambda$ in all
figures.}\label{Fig2}\vspace{-.8cm}
\end{figure}

First, we discuss the resolution of a circular aperture. The
source is located just behind the aperture at a subwavelength
distance and we consider the field at the far zone as depicted in
Fig. \ref{Fig1}. We assume that the aperture is inside a medium
which is linear, isotropic, homogenous, non-dispersive and
non-magnetic. Under these assumptions, all the components of the
electromagnetic field are uncoupled and the space dependent part
$U(\vec{r})$ satisfies the Helmholtz wave equation at each
source-free point:
 \begin{equation}\label{Helmholtz}
(\nabla^2+k^2)U(\vec{r})=0.
 \end{equation}
For simplicity, we assume index of refraction equal to 1 so that
$k=\omega/c$. For diffraction by planar screens, equation
(\ref{Helmholtz}) can be solved by following Kirchhoff in applying
Green's theorem,
% \begin{align}\label{Kirchoff}
% U(\vec{r},\vec{R})&=-\frac{iA}{2\lambda}\iint_S dS\frac{e^{ik_0|\vec{R}-\vec{\rho_0}|}}{|\vec{R}-\vec{\rho_0}|}
% \frac{e^{ik_0|\vec{r}-\vec{\rho_0}|}}{|\vec{r}-\vec{\rho_0}|}\left\{\frac{R_z}{|\vec{R}-\vec{\rho_0}|}\right. \nonumber \\
% &\!\!\!\!\!\!\!\!\!\!\!\!\!\!\!\! \left.
% \left(1-\frac{1}{ik_0|\vec{R}-\vec{\rho_0}|}\right)+\frac{r_z}{|\vec{r}-\vec{\rho_0}|}\left(1-\frac{1}{ik_0|\vec{r}-\vec{\rho_0}|}\right)\right\},
% \end{align}
%  \begin{align}\label{Kirchoff}
% U(\vec{r},\vec{R})&=-\frac{iA}{2\lambda}\iint_S dS\frac{e^{ik_0|\vec{R}-\vec{\rho_0}|}}{|\vec{R}-\vec{\rho_0}|}
% \frac{e^{ik_0|\vec{r}-\vec{\rho_0}|}}{|\vec{r}-\vec{\rho_0}|}\nonumber \\
% &\left\{\frac{R_z}{|\vec{R}-\vec{\rho_0}|}\left(1-\frac{1}{ik_0|\vec{R}-\vec{\rho_0}|}\right)\right.\nonumber\\
% &\left.+\frac{r_z}{|\vec{r}-\vec{\rho_0}|}\left(1-\frac{1}{ik_0|\vec{r}-\vec{\rho_0}|}\right)\right\},
% \end{align}
   \begin{align}\label{Kirchoff}
 U(\vec{r},\vec{R})&=-\frac{1}{4\pi}\iint_S dS\left\{\frac{\partial U}{\partial n}
 \left[\frac{e^{ik|\vec{r}-\vec{\rho}|}}{|\vec{r}-\vec{\rho}|}\right]\right.\nonumber \\
 &\left.-U\frac{\partial}{\partial n}\left[\frac{e^{ik|\vec{r}-\vec{\rho}|}}{|\vec{r}-\vec{\rho}|}\right]\right\},
 \end{align}
where the integral is taken over the plane of aperture and
$\partial/\partial n$ signifies a partial derivative in the
outward normal direction at each point on $S$. The vectors $\vec
r$ and $\vec R$ extends from the center of the aperture to the
source and the screen respectively and $\vec\rho$ scans the
aperture plane (see Fig. \ref{Fig1}). The field amplitude $U$ on
the aperture plane $S$ is given by the spherical wave
$\exp(ik|\vec{R}-\vec{\rho}|)/{|\vec{R}-\vec{\rho}|}$. If both the
source and the detection plane are located in the far zone one can
make use of the Fresnel and Fraunhofer approximations,
$|\vec{r}-\vec{\rho}|,|\vec{R}-\vec{\rho}|\gg k|\vec\rho|^2/2$,
%\begin{equation}\label{Fraunhofer}
%|\vec{r}-\vec{\rho}|,|\vec{R}-\vec{\rho}|\gg k|\vec\rho|^2/2,
%\end{equation}
and the integral given in Eq. (\ref{Kirchoff}) can be carried out
to obtain,
\begin{equation}\label{fraunhofer diffraction}
U(\vec{r},\vec{R})=-\frac{iAkae^{ik(r_z+R_z)}e^{i\frac{k}{2}\frac{r_{||}^2}{r_z}}}{R_zr_z}\frac{J_1\left(2\pi\xi
a\right)}{2\pi\xi},
\end{equation}
where A is a constant proportional to amplitude of the electric
field at unit distance from the source, $a$ is the radius of the
circular aperture, and $\xi=r_{||}/\lambda r_z$ is radius in the
spatial frequency domain. Here $r_{||}=\sqrt{r_x^2+r_y^2}$ is the
radial distance to the optical axis on the detection plane. The
intensity distribution associated with Eq. (\ref{fraunhofer
diffraction}) is referred to as the well known Airy pattern. It is
rotationally symmetric around the $z$-axis. Figure \ref{Fig2}$(a)$
shows a cross section of the Airy pattern for an aperture size
$a=2\lambda$. The position of the first minima, measured along the
radial axis, is given by $\Delta r=0.61\frac{\lambda}{a/r_z}$. The
width $\Delta r$ is also denoted as the \emph{Airy disk radius}.
The ability to resolve two apertures depends on the size of the
Airy disk. The narrower the width, $\Delta r$, is the better the
resolution will be. For far field detection, \emph{Airy disk
radius} can be on the order of wavelength only if the size of the
aperture is much larger than the wavelength itself. On the other
hand, an aperture having a size of one wavelength will produce an
Airy disk of radius $0.61r_z$, which is much larger than the
wavelength because $r_z\gg\lambda$. The effective size of the
diffraction pattern is seen to be inversely proportional to the
linear dimensions of the aperture. This is a direct consequence of
diffraction of waves and spreading comes from the uncertainty
relation $\Delta \rho_{||}\Delta k_{\rho_{||}}\geq 2\pi$. In
far-field optics, the upper bound for $\Delta k_{\rho_{||}}$ is
given by the wavenumber $k=2\pi/\lambda$ of the object medium
because we discard spatial frequencies associated with evanescent
wave components. In this case the resolution cannot be better than
$min[\Delta \rho_{||}]=\lambda/2$ and one can reconstruct a
low-pass filtered version of the aperture.
%If $\theta$ is the
%semiaperture angle from the center of the aperture axis to the
%first minimum of intensity, the minimum distance resolved is given
%by
%\begin{equation}\label{minima position}
%\Delta r=\frac{0.61\lambda_0}{\sin\theta}.
%\end{equation}

Now, let us consider the setup in which the dipole is located very
close to the aperture. In this case, Fraunhofer approximations are
not valid in the region between the dipole and the aperture. Then,
the integral given in Eq. (\ref{Kirchoff}) becomes,
   \begin{align}\label{near field integral}
 U(\vec{r},\vec{R})&=-\frac{\Phi}{r_z\lambda}\iint_S d\rho_xd\rho_ye^{i\frac {k}{r_z}\left (\rho_xr_x + \rho_yr_y \right)}\nonumber\\
 &\frac{e^{ik|\vec{R}-\vec{\rho}|}}{|\vec{R}-\vec{\rho}|}\frac{R_z}{|\vec{R}-\vec{\rho}|}\left(1-\frac{1}{ik|\vec{R}-\vec{\rho}|}\right),
 \end{align}
where $\Phi=\exp\{ikr_z+ik(r_x^2+r_y^2)/2r_z\}$ is a phase factor.
Our aim is to compare this result with Eq. (\ref{fraunhofer
diffraction}). To analyze the angular spectrum of the field one
can expand the spherical wave into an angular spectrum of plane
waves as given by Weyl's \cite{Weyl} integral,
\begin{equation}\label{weyl decomposition}
\frac{e^{ik|\vec{R}-\vec{\rho}|}}{|\vec{R}-\vec{\rho}|}\!\!=\!\!\frac{i}{2\pi}\!\!\iint_{-\infty}^{+\infty}
\!\!\frac{dk_xdk_y}{k_z}e^{-i\left[k_x(R_x-\rho_x)+k_y(R_y-\rho_y)+k_zR_z\right]},
\end{equation}
 where,
\begin{equation}
k_z = \left\{
\begin{array}{rl}\label{kz}
\sqrt{k^2-k_{||}^2} & \text{for } k_{||}^2\leq k^2,\\
i\sqrt{k_{||}^2-k^2} & \text{for } k_{||}^2> k^2,
\end{array} \right.
\end{equation}
$R_z=-\epsilon$ is the source-aperture distance on the $z$-axis
and $k_{||}^2=k_x^2+k_y^2$.
%Each plane wave in the spectrum of Eq. (\ref{weyl
%decomposition}) is a conventional homogenous wave if
%$k_x^2+k_y^2\leq k^2$ or an evanescent plane wave if $k_x^2+k_y^2>
%k^2$.
By substitution from Eq. (\ref{kz}) and Eq. (\ref{weyl
decomposition}) into Eq. (\ref{near field integral}) it can be
verified that the angular spectrum of $ U(\vec{r},\vec{R})$ has
the following transverse wave vector component
\begin{equation}\label{wave vector}
\vec{\kappa}_{||}=k\frac{\vec{r}_{||}}{r_z}+\vec{k}_{||}.
\end{equation}
%Eq. (\ref{wave vector}) shows that angular spectrum is now the
%main spatial frequency, $k$, multiplied by the numerical aperture
%($r_{||}/r_z$)
This contains the source wave vector multiplied by the numerical
aperture ($r_{||}/r_z$) and the lateral wave vector,
$\vec{k}_{||}$, of the dipole. Therefore the spatial frequency
exceeds that of Eq. (\ref{fraunhofer diffraction}) by $k_{||}$. In
principle, there is no upper limit for $k_{||}$ because a point
source in space is characterized by an infinite spectrum of
spatial frequencies $k_x$, $k_y$. A spherical wave is well
localized in position but its momentum has a direction that is
totally uncertain. For large $\epsilon$, high frequency components
are filtered out and we get, $\kappa_{||}\leq k$, i.e. the usual
diffraction limit.
% evanescent plane waves attenuate
%exponentially with increasing $|R_z|$. This is the basis for the
%usual argument that evanescent plane waves can be neglected
%sufficiently far away from the $z=0$ plane.
On the other hand, for $\epsilon<\lambda$, the contribution coming
from the higher spatial frequency modes, ($k_{||}^2>k^2$) can not
be ignored. It is possible to collect higher spatial frequencies
partially by locating the dipole closer to the aperture. Note that
even for $k_{||}^2\equiv k_x^2+k_y^2>k^2$, the wave in Eq.
(\ref{near field integral}) is propagating because of the phase
factor $\Phi$.
%The contribution associated with the
%higher spatial frequency components are then carried out to the
%far zone by the integral given in Eq. (\ref{near field integral}).

To see the effect of the near field of the dipole we numerically
calculate the intensity associated with the integral given in Eq.
(\ref{near field integral}). Figures
\ref{Fig2}$(a)$-\ref{Fig2}$(d)$ represent the intensity of
diffracted field for a given radius of the aperture and for
different values of dipole-aperture distance, $\epsilon$. The
bandwidth of spatial frequencies increases as $\epsilon$
decreases, because the contribution coming from larger $k_{||}$
values increases as $\epsilon$ gets smaller in the integral given
by Eq. (\ref{weyl decomposition}).
Comparing Fig. \ref{Fig2}$(a)$ to Figs.
\ref{Fig2}$(b)-\ref{Fig2}(d)$ we can understand how the near field
of the dipole increases $k_{||}$ and changes the resolution of the
object. Higher spatial frequencies, $k_{||}^2>k^2$, which are
totally lost in classical imaging (see Fig. \ref{Fig2}$(a)$) are
partially converted into propagating waves. What we are showing is
in a sense inverse of what one does in Fresnel diffraction as we
keep the source distance much smaller than a wavelength. However
the observation plane is in far field. The precise mathematical
connection between our results and fresnel diffraction needs to be
examined further using the rigorous wave equations.

\begin{figure}[h]
\centering {\includegraphics[width=405pt]{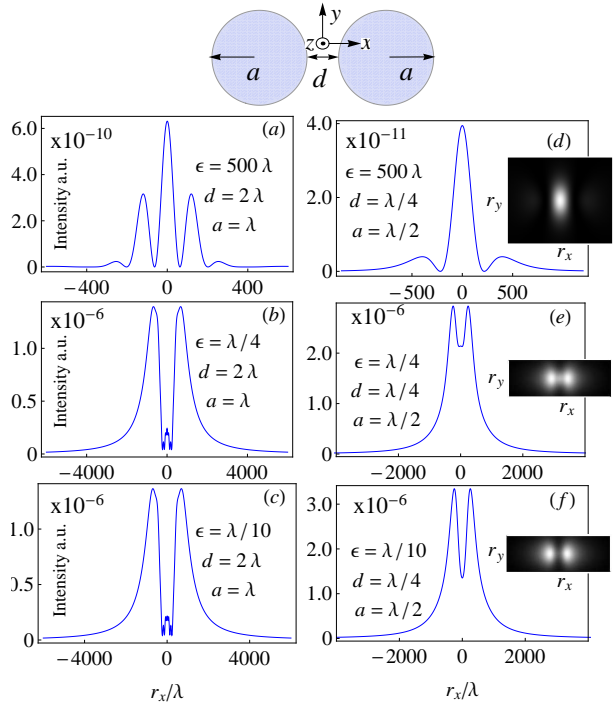}}\vspace{-.8cm}
\caption{Computed far field images of two separated circular
apertures. $r_z$ is $500\lambda$ in all figures. The insets in
$(d)$, $(e)$ and $(f)$ show the intensities in two
dimensions.}\label{Fig3}\vspace{-.75cm}
\end{figure}

The far field resolution enhancement provided by the evanescent
component of the dipole can also be considered for the diffraction
by two closely located apertures. We calculate the far field using
the same scalar diffraction integral in Eq. (\ref{near field
integral}). This time the integration surface is chosen to be two
circular holes as illustrated in Fig. \ref{Fig3}. We choose the
radii of the apertures to be $\lambda$ and $\lambda/2$. We
consider two different separation distances, $d=2\lambda$ and
$d=\lambda/4$, measured from the rim. First we calculate the
patterns for the source located in the far zone for comparison
(see Figs. \ref{Fig3} ($a$) and \ref{Fig3} ($d$)). The images look
like an interference pattern from two slits and the objects are
not resolved. As the source gets closer in the subwavelength
range, the objects appear as two separated spots and they are
resolved (see Figs. \ref{Fig3} $(b), (c)$ and \ref{Fig3} $(e),
(f)$).

To summarize, we have described how the high spatial frequencies
of a localized scalar field can be utilized to enhance the
resolution of objects in the far zone. We use the scalar
diffraction theory and the method of angular spectrum
representation to compute the images numerically. The principle of
enhanced image resolution is explained by analyzing the transverse
wave vector component of the the diffracted field in the object
plane. We also demonstrate how partially collected evanescent
waves by the scalar dipole can resolve two closely located
objects. The principle of resolution enhancement by using
evanescent waves in this paper do not require a scanning probe or
near-field detection schemes. Therefore, the far-field microscopy
techniques can be employed for the image reconstruction.

A.K. is grateful to the Scientific and Technological Research
Council of Turkey (T\"{U}B\.{I}TAK) grant no 110T321 for
supporting this research.\vspace{-1.0cm}

\end{document}